\documentclass[prb,aps,twocolumn,floatfix,showpacs,letter]{revtex4}
\usepackage{bm}
\usepackage{graphicx}

\begin{document}
\title{Ginzburg-Landau theory of vortices in a multi-gap superconductor}
\author{M. E. Zhitomirsky and  V.-H. Dao}
\affiliation{Commissariat \'a l'Energie Atomique, DMS/DRFMC/SPSMS, 
17 avenue des Martyrs, 38054 Grenoble, Cedex 9 France}
\date{\today}

\begin{abstract}
The Ginzburg-Landau functional for a two-gap superconductor
is derived within the weak-coupling BCS model. The two-gap Ginzburg-Landau 
theory is, then, applied to investigate various magnetic properties 
of MgB$_2$ including an upturn temperature dependence of the
transverse upper critical field and a core structure of an isolated vortex.
Orientation of vortex lattice relative to crystallographic axes
is studied for magnetic fields parallel to the $c$-axis.
A peculiar $30^\circ$-rotation of the vortex lattice with increasing strength
of an applied field observed by neutron scattering is attributed 
to the multi-gap nature of superconductivity in MgB$_2$.
\end{abstract}

\maketitle

\section{Introduction}

Superconductivity in MgB$_2$ discovered a few years ago\cite{akimitsu}
has attracted a lot of interest both from fundamental and technological
points of view.\cite{canfield}
Unique physical properties of MgB$_2$ include $T_c=39$~K, the highest
among $s$-wave phonon mediated superconductors, and the presence
of two gaps $\Delta_1\approx 7$~meV and $\Delta_2\approx 2.5$~meV
evidenced by the scanning tunneling\cite{roditchev,ivarone} and 
the point contact\cite{szabo,schmidt} spectroscopies and by the heat capacity
measurements.\cite{wang,bouquet1,yang,bouquet2} The latter property brings
back the concept of a multi-gap superconductivity\cite{suhl,moskalenko}
formulated more than forty years ago for materials with large disparity
of the electron-phonon interaction for different pieces of the Fermi
surface.

Theoretical understanding of normal and superconducting properties of 
MgB$_2$ has been advanced in the direction of first-principle calculations 
of the electronic band structure and the electron-phonon interaction, 
which identified two distinct groups of bands with large and small 
superconducting gaps.\cite{kortus,belaschenko,an,kong,liu,choi,mazant}
Quantitative analysis of various thermodynamic and transport properties
in the superconducting state of MgB$_2$ was made in the framework of
the two-band BCS model. 
\cite{golubov,brinkman,mazin,nakai,miranovic,dahm,gurevich,golubov2,koshelev}
An outside observer would notice, however, a certain lack of effective 
Ginzburg-Landau or London type theories applied to MgB$_2$. This fact 
is explained of by quantitative essence of the discussed problems, 
though effective theories can often give a simpler insight. Besides, 
new experiments constantly raise different types of questions. For example, 
recent neutron diffraction study in the mixed state of MgB$_2$ has found 
a strange $30^\circ$-reorientation of the vortex lattice with increasing 
strength of a magnetic field applied along the $c$-axis.\cite{cubitt2}
Such a transition represents a marked qualitative departure from the 
well-known behavior of the Abrikosov vortex lattice in single-gap 
type-II superconductors. Nature and origin of phase transitions in 
the vortex lattice are most straightforwardly addressed by the
Ginzburg-Landau theory. 

In the present work we first derive the appropriate Ginzburg-Landau 
functional for a two-gap superconductor from the microscopic BCS model.
We, then, investigate various magnetic properties of MgB$_2$ using
the Ginzburg-Landau theory. Our main results include demonstration
of the upward curvature of $H_{c2}(T)$ for transverse magnetic fields,
investigation of the vortex core structure, and explanation
of the reorientational transition in the vortex lattice.
The paper is organized as follows. Section 2 describes
the two-band BCS model and discusses the fit of experimental data
on the temperature dependence of the specific heat.
Section 3 is devoted to derivation of the Ginzburg-Landau functional
for a two-gap weak-coupling superconductor. In Section~4
we discuss various magnetic properties including the upper critical
field and the structure of an isolated vortex.
Section 5 considers the general problem of an orientation of the vortex
lattice in a hexagonal superconductor in magnetic field
applied parallel to the $c$-axis and, then, demonstrates how
the multi-gap nature of superconductivity in MgB$_2$
determines a reorientational transition in the mixed state.

\section{The two-band BCS model}

\subsection{General theory}

In this subsection we briefly summarize the thermodynamics of an 
$s$-wave two-gap superconductor with the aim to extract subsequently 
microscopic parameters of the model from available experimental data 
for MgB$_2$. We write the pairing interaction as
\begin{equation}
\hat{V}_{\rm BCS}= -\sum_{n,n'} g_{nn'} \int dx
\Psi^\dagger_{n\uparrow}(x) \Psi^\dagger_{n\downarrow}(x)
\Psi_{n'\downarrow}(x)\Psi_{n'\uparrow}(x) \ ,
\label{bcs}
\end{equation}
where $n=1,2$ is the band index. A real space representation (\ref{bcs}) is
obtained from a general momentum-space form of the 
model\cite{suhl,moskalenko}
under assumption of weak momentum dependence of the scattering amplitudes
$g_{nn'}$. We also assume that the  active band has the strongest pairing
interaction $g_{11}=g_1$ compared to interaction in the passive band
$g_{22}=g_2$ and to interband scattering of the Cooper pairs
$g_{12}=g_{21}=g_3$. Defining two gap functions
\begin{equation}
\Delta_n(x) =  - \sum_{n'} g_{nn'} \langle\Psi_{n'\downarrow}(x)
\Psi_{n'\uparrow}(x)\rangle
\label{gapeq}
\end{equation}
the total Hamiltonian is transformed to the mean-field form
\begin{eqnarray}
\hat{H}_{\rm MF} & = & E_{\rm const} + \sum_n \int dx
\Bigl[\Psi^\dagger_{n\sigma}(x)\hat{h}(x)\Psi_{n\sigma}(x)  \nonumber \\
& & \mbox{} + \Delta_n(x) \Psi^\dagger_{n\uparrow}(x)
\Psi^\dagger_{n\downarrow}(x) +
h.c. \Bigr] \ ,
\label{MF}
\end{eqnarray}
$\hat{h}(x)$ being a single-particle Hamiltonian of the normal metal. 
The constant term is a quadratic form of anomalous averages
$\langle\Psi_{n\downarrow}(x)\Psi_{n\uparrow}(x)\rangle$.
Using Eq.~(\ref{gapeq}) it can be expressed via the gap functions
\begin{equation}
E_{\rm const}\! =\! \frac{1}{G} \int dx \bigl[g_2|\Delta_1|^2\!+\!
g_1|\Delta_2|^2-g_3( \Delta^*_1\Delta_2\!+\!\Delta^*_2\Delta_1 )\bigr]
\label{const}
\end{equation}
with $G=\det\{g_{nn'}\}= g_1g_2-g_3^2$. The above expression has
to be modified for $G=0$. In this case the two equations
(\ref{gapeq}) are linearly dependent. As a result, the ratio of the two
gaps is the same for all temperatures and magnetic fields
$\Delta_2(x)/\Delta_1(x)=g_3/g_1$, while the constant term reduces
to $E_{\rm const}=\int dx\; |\Delta_1|^2/g_1$.

The standard Gorkov's technique can then be applied to derive the
Green's functions and energy spectra in uniform and nonuniform
states with and without impurities. In a clean superconductor in
zero magnetic field the two superconducting gaps are related via
the self-consistent gap equations
\begin{equation}
\Delta_n = \sum_{n'} \lambda_{nn'}\Delta_{n'} \int_0^{\omega_D}
\frac{d\varepsilon} {\sqrt{\varepsilon^2+\Delta_{n'}^2}}\tanh
\frac{\sqrt{\varepsilon^2+\Delta_{n'}^2}} {2T}
\label{gapeq2}
\end{equation}
with dimensionless coupling constants $\lambda_{nn'}=g_{nn'}
N_{n'}$, $N_{n}$ being the density of states at the Fermi level
for each band. The transition temperature is given by
$T_c = (2\omega_D e^C/\pi) e^{-1/\lambda}$, where $\omega_D$
is the Debay frequency, $C$ is the Euler constant and $\lambda$
is the largest eigenvalue of the matrix $\lambda_{nn'}$:
\begin{equation}
\lambda = 
(\lambda_{11}+\lambda_{22})/2+\sqrt{(\lambda_{11}-\lambda_{22})^2/4
+\lambda_{12}\lambda_{21}} \ . \nonumber
\end{equation}
Since $\lambda>\lambda_{11}$, the interband scattering always increases
the superconducting transition temperature compared to an instability
in a single-band case. The ratio of the two gaps at $T=T_c$ is
$\Delta_2/\Delta_1 = \lambda_{21}/(\lambda-\lambda_{22})$. At zero
temperature the gap equations (\ref{gapeq2}) are reduced to
\begin{equation}
\Delta_n = \sum_{n'} \lambda_{nn'} \Delta_{n'}
\ln\frac{2\omega_D}{\Delta_{n'}} \ .
\end{equation}
By substituting $\Delta_n = 2\omega_D r_n e^{-1/\lambda}$ the above
equation is transformed to
\begin{equation}
r_n = \sum_{n'} \lambda_{nn'} r_{n'} \biggl(\frac{1}{\lambda} -\ln
r_{n'}\biggr) \ .
\label{ratio0}
\end{equation}
For $1/\lambda \gg \ln r_n$, one can neglect logarithms on the right hand
side and obtain for the ratio of the two gaps the same equation as at
$T=T_c$ implying that $\Delta_2/\Delta_1$ is temperature
independent.\cite{zaitsev} This approximation is valid only for
$r_n\simeq 1$, {\it i.e.\/}, if all the coupling constants
$\lambda_{nn'}$ have the same order of magnitude. (For $g_3^2=g_1g_2$
the above property is an exact one: the gap ratio
does not change neither with temperature nor in magnetic field.)
However, for $g_3\ll g_2<g_1$, the passive gap $\Delta_2$ is
significantly smaller than the active gap $\Delta_1$ and $r_2\ll 1$
so that the corresponding logarithm cannot be neglected. It follows
from Eq.~(\ref{ratio0}) that the ratio $\Delta_2/\Delta_1$
increases between $T=T_c$ and $T=0$ for small $g_3$.
Such variations become more pronounced in superconductors
with larger values of $\lambda$, which are away from the extreme
weak-coupling limit $\lambda\ll 1$. Ab-initio calculations indicate that
MgB$_2$ has an intermediate strength of the electron-phonon coupling
with $\lambda_{12(21)}\ll \lambda_{11}\alt 1$, making this superconductor
an ideal system to observe effects related to variations of the ratio of
two gaps.

The jump in the specific heat at the superconducting transition
can be expressed analytically as\cite{moskalenko,zaitsev,mishonov}
\begin{equation}
\frac{\Delta C}{C}=\frac{12}{7\zeta(3)}
\frac{(N_1\Delta_1^2+N_2\Delta_2^2)^2}
{(N_1+N_2)(N_1\Delta_1^4+N_2\Delta_2^4)} \ ,
\end{equation}
where the limit $T\rightarrow T_c$ has to be taken for the ratio of
the two gaps. The specific heat jump is always smaller than
the single-band BCS result $ \Delta C/C=12/7\zeta(3)\approx 1.43$,
unless $\Delta_1=\Delta_2$.

\subsection{Fit to experimental data}

One of the striking experimental evidences of the double-gap
behavior in MgB$_2$ is an unusual temperature dependence of
the specific heat with a shoulder-type anomaly around
$0.25T_c$.\cite{wang,bouquet1,yang,bouquet2} We use here the
multi-band BCS theory to fit the experimental data
for $C(T)$. The Fermi surface in MgB$_2$ consists of four sheets:
two nearly cylindrical hole sheets arising from quasi two-dimensional
$p_{x,y}$ boron bands and two sheets from three-dimensional $p_z$
bonding and antibonding bands.\cite{FS,kortus} The electronic
structure of MgB$_2$ is now well understood from a number of
density-functional studies,\cite{kortus,belaschenko,an,kong,liu,choi,mazant}
which generally agree with each other, though differ in certain
details. Specifically, we choose as a reference the work of Kong
{\it et al\/}.,\cite{kong} where the tight-binding fits for all Fermi
surface sheets in MgB$_2$ are provided. Using these fits we have calculated
various Fermi surface averages for each band. The density of states
at the Fermi level is $N(0) = 0.41$~states/eV/cell/spin, which includes
$N_\sigma(0)=0.16=0.049+0.111$~states/eV/cell/spin in light and heavy
$\sigma$-bands and $N_\pi(0)=0.25=0.124+0.126$~states/eV/cell/spin in
the two $\pi$-bands. Note, that the obtained $N_\pi(0)$ is somewhat
larger than the number $0.205$ cited by Kong {\it et al\/}.,\cite{kong}
while the results for the $\sigma$-bands agree. Because of a strong
mismatch in the electron-phonon coupling between two group of
bands,\cite{an,kong,liu,choi} the two $\sigma$-bands can be represented
as a single active band, which has $N_1=0.4N(0)$ of the total density
of states and drives superconducting instability, whereas a combined
$\pi$-band contributes $N_2=0.6N(0)$ to the total density of states
and plays a passive role in the superconducting instability. The
above numbers are consistent with $N_1=0.45N(0)$ and $N_1=0.42N(0)$
for the partial density of states of the of the electrons in the
$\sigma$-bands obtained in the other studies.\cite{liu,belaschenko}

\begin{figure}[t]
\begin{center}
\includegraphics[width=0.95\columnwidth]{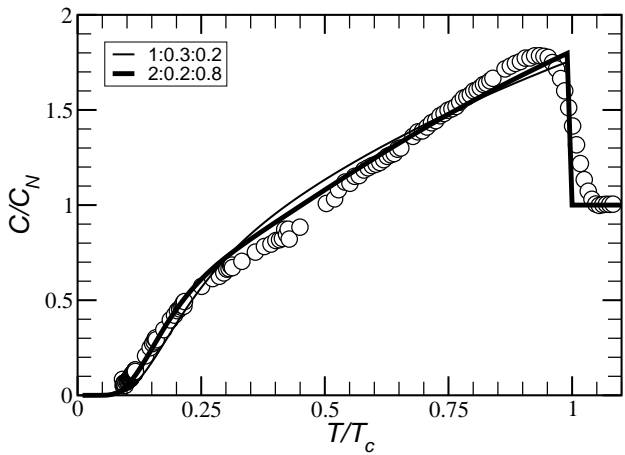}
\end{center}
\caption{
Theoretical dependence of the specific heat in the two-band BCS
model. Numbers for each curve indicate
values of $g_1$, $g_2$, and $g_3$ ($N_1=0.4$, $N_2=0.6$).
Open circles are the experimental data.$^{7,10}$}
\label{Heat}
\end{figure}

The gap equations (\ref{gapeq2}) have been solved self-consistently for 
$N_2/N_1=1.5$ and various values of coupling constants.
The specific heat is calculated from
\begin{equation}
C(T) = \sum_{n k} E_{nk} \frac{d n_F(E_{nk})}{dT} \ ,
\end{equation}
where $E_{nk}=\sqrt{\varepsilon_k+\Delta_n^2}$ is a quasiparticle energy
for each band and $n_F(\varepsilon)$ is the Fermi distribution. 
Figure~\ref{Heat} shows two theoretical fits to the experimental data 
of Geneva group\cite{wang,bouquet2} using a weak $g_1N_1=0.4$ and 
a moderate $g_1N_1=0.8$ strength of the
coupling constant in the active band. Constants $g_2$ and $g_3$ have
been varied to get the best fits. In the first case the gap ratio changes
in the range $\Delta_1/\Delta_2=$3.--2.5 between $T=T_c$ and $T=0$,
while in the second case $\Delta_1/\Delta_2\simeq 2.7$. Both theoretical
curves reproduce quite well the qualitative behavior of $C(T)$. Somewhat
better fits can be obtained by increasing the partial density of states
in the $\sigma$-band. Quantitative discrepancies between various
theoretical fits and the experimental data are, however, less significant
than differences between different samples. \cite{bouquet2} We, therefore,
conclude that though the specific heat data clearly agree with the
two-gap superconducting model in the regime of weak interband interaction,
a unique identification of coupling constants is not possible from
available data.

\section{the Ginzburg-Landau functional}

We use the microscopic theory formulated in the previous section
to derive the Ginzburg-Landau functional of a two-gap superconductor. In
the vicinity of $T_c$ the anomalous terms in the mean-field Hamiltonian
(\ref{MF}) are treated as a perturbation $V_a$. Then, the thermodynamic
potential of the superconducting state is expressed as
\begin{equation}
\Omega_s = E_{\rm const}-\frac{1}{\beta}\ln\left\langle
T_\tau\exp\Bigl[-\int_0^\beta V_a(\tau) d \tau \Bigr]\right\rangle \ ,
\label{perturb}
\end{equation}
where $\beta=1/T$. Expansion of Eq.~(\ref{perturb}) in powers of
$V_a$ yields the Ginzburg-Landau functional.
Since the normal-state Green's functions are diagonal in the band index,
the Wick's decoupling of $V_a$ in $\Omega_s$ does not produce any
mixing terms between the gaps.
As a result, the weak-coupling Ginzburg-Landau functional
has a single Josephson-type interaction term:
\begin{eqnarray}
&& F_{GL} = \int dx \Bigl[ \alpha_1 |\Delta_1|^2  + \alpha_2 |\Delta_2|^2
- \gamma (\Delta_1^*\Delta_2 + \Delta_2^*\Delta_1)  \nonumber \\
&& \mbox{}\!+\textstyle\frac{1}{2}\beta_1|\Delta_1|^4\!+
\textstyle\frac{1}{2}\beta_2|\Delta_2|^4\!+ K_{1i}|\nabla_i\Delta_1|^2\!+
K_{2i} |\nabla_i\Delta_2|^2\Bigr], \nonumber \\
&& \nabla_i = \partial_i + i\frac{2\pi}{\Phi_0} A_i ,\
\alpha_{1,2} = \frac{g_{2,1}}{G}- N_{1,2}\ln\frac{2\omega_D e^C}{\pi T},
\label{GL}\\
&&\beta_n = \frac{7\zeta(3)N_n}{16\pi^2 T_c^2}, \ \ \gamma = \frac{g_3}{G},\ 
\
K_{ni} = \frac{7\zeta(3)N_n}{16\pi^2 T_c^2}\langle v_{Fni}^2\rangle \ ,
\nonumber
\end{eqnarray}
$\Phi_0$ being the flux quantum. For $\gamma>0$, the interaction term favors
the same phase for the two gaps. For $\gamma<0$, if, {\em e.g.\/},
the Coulomb interactions dominate the interband scattering of the Cooper
pairs and $g_3<0$, the smaller gap acquires a $\pi$-shift
relative to the larger gap.\cite{mazgol,imada}

The gradient term coefficients depend
in a standard way on the averages of Fermi velocities ${\bf v}_{Fn}$
over various sheets of the Fermi surface. Numerical integration
of the tight-binding fits\cite{kong} yields the following results:
for the $\sigma$-band
$\langle v_{Fx}^2\rangle = 2.13$ (3.55, 1.51)  and
$\langle v_{Fz}^2\rangle = 0.05$ (0.05, 0.05); for the $\pi$-band
$\langle v_{Fx}^2\rangle = 1.51$ (1.47, 1.55) and
$\langle v_{Fz}^2\rangle = 2.96$ (2.81, 3.10) in units of
$10^{15}$~cm$^2$/s$^2$, numbers in parentheses correspond to each
of the constituent bands.
The effective masses of the quasi two-dimensional $\sigma$-band exhibit
a factor of 40 anisotropy between in-plane and out of plane
directions. In contrast, the three-dimensional $\pi$-band has a somewhat
smaller mass along the $c$-axis. Using $N_2/N_1=1.5$ we find
that the in-plane gradient constants for the two bands
are practically the same $K_{2\perp}/K_{1\perp}\approx 1.06$,
while the $c$-axis constants differ by almost two orders
of magnitude $K_{2z}/K_{1z}\approx 90$.

A very simple form of the two-gap weak-coupling Ginzburg-Landau
functional is somewhat unexpected. On general symmetry grounds,
there are possible various types of interaction in
quartic and gradient terms between two superconducting condensates
of the same symmetry, which have been considered in the 
literature.\cite{sigrist,betouras,vinokur} The above form of the 
Ginzburg-Landau functional is, nevertheless, a straightforward extension 
of the well-known result for unconventional superconductors. For example, 
the quartic term for a momentum-dependent gap is $|\Delta({\bf k})|^4$ in 
the weak-coupling approximation.\cite{gorkov,leggett}
In the two-band model $\Delta({\bf k})$ assumes a step-like dependence
between different pieces of the Fermi surface, which immediately leads to
the expression (\ref{GL}).

The Ginzburg-Landau equations for the two-gap superconductor,
which are identical to those obtained from Eq.~(\ref{GL}), have
been first derived by an expansion of the gap equations
in powers of $\Delta$.\cite{zaitsev} Recently, a
similar calculation has been done for a dirty superconductor,
with only {\em intraband} impurity scattering
and the corresponding form of the Ginzburg-Landau functional
has been guessed, though with incorrect sign of the coupling
term.\cite{gurevich}
Here, we have directly derived the free energy of the two-gap
superconductor. The derivation can be straightforwardly generalized
to obtain, e.g. higher-order gradient terms, which are needed
to find an orientation of the vortex lattice relative to crystal axis
(see below). We also note that strong-coupling effects, e.g., dependence of
the pairing interactions on the gap amplitudes, will produce
other, generally weaker, mixing terms of the fourth order in $\Delta$.
The {\em interband} scattering by impurities can generate a mixing gradient
term as well.

Finally, for $G=(g_1g_2-g_3^2)<0$ a number of spurious features appears
in the theory: the matrix $\lambda_{nn'}$ and the quadratic form
(\ref{const}) acquire negative eigenvalues, while a formal minimization
of the Ginzburg-Landau functional (\ref{GL}) leads to an unphysical
solution at high temperatures. Sign of $\Delta_2/\Delta_1$ for such a
solution is opposite to the sign of $g_3$. The origin of this
ill-behavior lies in the approximation of positive integrals on the
right-hand side of Eq.~(\ref{gapeq2}) by logarithms, which can become
negative. Therefore, negative eigenvalues of $\lambda_{nn'}$ and
$E_{\rm const}$ yield no physical solution similar to the case when the
BCS theory is applied to the Fermi gas with repulsion. The consequence
for the Ginzburg-Landau theory (\ref{GL}) is that one should keep
the correct sign of $\Delta_2/\Delta_1$ and use the Ginzburg-Landau
equations, {\em i.e.\/}, look for a saddle-point solution rather
than seeking for an absolute minimum.

\section{The two-gap Ginzburg-Landau theory}

In order to discuss various properties of a two-gap superconductor
in the framework of the Ginzburg-Landau theory we write
$\alpha_1 = - a_1 t$ with $a_1=N_1$, $t = \ln(T_1/T)\approx (1-T/T_1)$
and $T_1= (2\omega_D e^C/\pi) e^{-g_2/GN_1}$ for the first active
band and $\alpha_2 = \alpha_{20} - a_2t$ with $a_2=N_2$, $\alpha_{20} =
(\lambda_{11}-\lambda_{22})/GN_1$ for the passive band.

\subsection{Zero magnetic field}

For completeness, we briefly mention here the behavior in zero
magnetic field. The transition temperature is found from
diagonalization of the quadratic form in Eq.~(\ref{GL}):
\begin{equation}
t_c = \frac{\alpha_{20}}{2a_2} - \sqrt{\frac{\alpha_{20}^2}{4a_2^2} +
\frac{\gamma^2}{a_1a_2}} \ .
\end{equation}
For small $\gamma$, one finds $t_c\approx - \gamma^2/(a_1\alpha_{20})$.
Negative sign of $t_c$ means that the superconducting transition takes
place above $T_1$, which is an intrinsic temperature of superconducting
instability in the first band. The ratio of the two gaps
$\rho=\Delta_2/\Delta_1 = \gamma/(\alpha_{20}-a_2t_c)$. Below the
transition temperature, the gap ratio $\rho$ obeys
\begin{equation}
\alpha_2\rho-\gamma+\frac{\beta_2}{\beta_1}\rho^3(a_1t+\gamma\rho)=0\ .
\label{ratio}
\end{equation}
For small $\gamma$, one can approximate $\rho\approx\gamma/\alpha_2$
and due to a decrease of $\alpha_2$ with temperature, small to
large gap ratio $\rho$ increases away from $t_c$.

\subsection{The upper critical field}

\subsubsection{Magnetic field parallel to the $c$-axis}

Due to the rotational symmetry about the $c$-axis, the gradient
terms in the $a$-$b$ plane are isotropic and can be described by
two constants $K_{n\perp}\equiv K_n$. The linearized Ginzburg-Landau 
equations describe a system of two coupled oscillators and have a 
solution in the form $\Delta_1 = c_0 f_0(x)$ and
$\Delta_2 = d_0 f_0(x)$, where $f_0(x)$ is a state on the zeroth
Landau level. The upper critical field is given by $H_{c2}=
h_{c2}\Phi_0/2\pi$
\begin{equation}
h_{c2} = \frac{a_1 t}{2K_1} - \frac{\alpha_2}{2K_2}
+  \sqrt{\left(\frac{a_1 t}{2K_1}+\frac{\alpha_2}{2K_2}\right)^2
+\frac{\gamma^2}{K_1K_2}}
\label{hc2_iso}
\end{equation}
The ratio of the two gaps $\rho=d_0/c_0$ along the upper critical
line is
\begin{equation}
\rho = \frac{\gamma}{\alpha_2+K_2 h_{c2}} \ .
\label{ratioH}
\end{equation}
The above expression indicates that an applied magnetic field
generally tends to suppress a smaller gap. Whether this effect
overcomes an opposite tendency to an increase of $\Delta_2/\Delta_1$
due to a decrease of $\alpha_2$ with temperature depends
on the gradient term constants. For example,
in the limit $\gamma \ll\alpha_{20}$ we find from (\ref{ratioH})
$\rho\approx \gamma/[\alpha_{20}-(a_2 -a_1 K_2/K_1)t]$.
If $K_2$ is significantly larger than $K_1$, while
$a_2\simeq a_1$, the smaller gap is quickly suppressed
along the upper critical field line. However, for MgB$_2$ one finds
$K_2/K_1\approx 1$ for in-plane gradient terms. Therefore,
the ratio $\Delta_2/\Delta_1$ continues to grow along $H_{c2}(T)$,
though slower than in zero field.

\begin{figure}[t]
\begin{center}
\includegraphics[width=0.9\columnwidth]{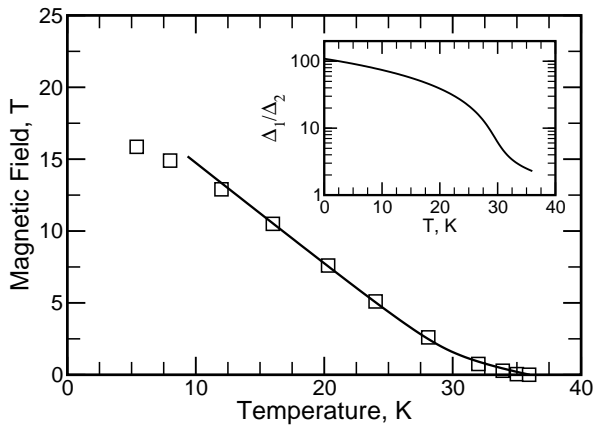}
\end{center}
\caption{
Temperature dependence of the upper critical field in MgB$_2$
for magnetic field in the basal plane. Solid line: the two-gap
Ginzburg-Landau theory with parameters given in the text,
squares: experimental data by Lyard {\it et al\/}.$^{40}$
The inset shows variation of the gap ratio along
the $H_{c2}(T)$ line for the same set of parameters.}
\label{Hc2}
\end{figure}

\subsubsection{Transverse magnetic field}

We assume that ${\bf H}\parallel \hat{\bf y}$ and consider a homogeneous
superconducting state along the field direction. The gradient terms in
two transverse directions $\hat{\bf x}$ and $\hat{\bf z}$ have different
stiffness constants $K_n$ and $K_{nz}$, respectively. In a single band
case, rescaling $x \rightarrow x(K_x/K_z)^{1/4}$ and
$z \rightarrow z(K_z/K_x)^{1/4}$ allows to reduce an anisotropic problem
to the isotropic one in rescaled coordinates. A multi-gap superconductor
has several different ratios $K_n/K_{nz}$ and the above rescaling
procedure does not work. In other words, coupled harmonic oscillators 
described by the linearized Ginzburg-Landau equations have different
resonance frequencies. To solve this problem we follow a variational approach, 
which is known to give a good accuracy in similar cases. The vector potential
is chosen in the Landau gauge ${\bf A} = (Hz,0,0)$ and we look for a
solution in the form
\begin{equation}
\left(\begin{array}{c}
\Delta_1 \\
\Delta_2 \end{array}\right)=\left(\frac{\lambda}{\pi}\right)^{1/4}
e^{-\lambda z^2/2}\left(\begin{array}{c} c \\ d \end{array}\right) ,
\end{equation}
where $\lambda$, $c$, and $d$ are variational parameters. After spatial
integration and substitution $\lambda=h/\mu$, $h=2\pi H/\Phi_0$, the
quadratic terms in the Ginzburg-Landau functional become
\begin{eqnarray}
F_2 & = & (-a_1 t + h\tilde{K}_1)|c|^2 + (\alpha_2 + h\tilde{K}_2)|d|^2  \\
& & - \gamma(c^* d+ d^* c)\ , \ \ \ \tilde{K}_n = \textstyle\frac{1}{2}
(K_n \mu + K_{nz}/\mu)
\nonumber
\end{eqnarray}
The determinant of the quadratic form vanishes at the transition into
superconducting state. Transition field is given by the same expression
as in the isotropic case (\ref{hc2_iso}), where $K_n$ have to be
replaced with $\tilde{K}_n$. The upper critical field is, then, obtained
from maximizing the corresponding expression with respect to the
variational parameter $\mu$. In general, maximization procedure has to
be done numerically. Analytic expressions are possible in two temperature 
regimes. At low temperatures $t\gg |t_c|$, the upper critical field is 
entirely determined by the active band and
\begin{equation}
h_{c2} = \frac{a_1t}{\sqrt{K_1K_{1z}}} \ .
\label{hc2low}
\end{equation}
In the vicinity of $T_c$, an external magnetic field has a small 
effect on the gap ratio $\rho=d/c\approx \gamma/\alpha_{20}$
and an effective single-gap Ginzburg-Landau theory can be applied.
The upper critical field is given by a combination of the gradient 
constants $K_{ni}$ weighted according to the gap amplitudes:
\begin{equation}
h_{c2} = \frac{a_1(t-t_c)}{\sqrt{(K_1+\rho^2 K_2)(K_{1z}+\rho^2K_{2z})}} \ .
\label{hc2high}
\end{equation}
Since, in MgB$_2$ one has $K_{1z}\simeq 0.01K_{2z}$ and $\rho^2\simeq 0.1$,
the slope of the upper critical field near $T_c$ is determined by an 
effective gradient constant $K_z^{\rm eff} \approx \rho^2K_{2z}>K_{1z}$ 
(while $(K_1 +\rho^2K_2)\approx K_1$). Thus, the upper critical field line 
$H_{c2}(T)$ shows a marked upturn curvature between the two regimes 
(\ref{hc2high}) and (\ref{hc2low}). Such a temperature behavior
has been recently addressed in a number of theoretical works 
based on various forms of the two-band BCS 
theory.\cite{miranovic,dahm,gurevich,golubov2}
We suggest here a simpler description of the above effect within
the two-gap Ginzburg-Landau theory.

Finally, we compare the Ginzburg-Landau theory
with the experimental data on the temperature dependence
of the upper critical field for magnetic field parallel
to the basal plane.\cite{samuely} We choose ratios of the
gradient term constants and the densities of states in accordance with
the band structure calculations\cite{kong} and change parameters
$\gamma$ and $\alpha_{20}$, which are known less accurately,
to fit the experimental data. The best fit shown in Fig.~\ref{Hc2}
is obtained for $\alpha_{20}/a_1=0.65$ and $\gamma/a_1=0.4$.
The prominent upward curvature of $H_{c2}(T)$ takes place between
$t_c=-0.18$ ($T_c=36$~K) and $t\simeq 0.2$ ($T=26$~K), i.e. well
within the range of validity of the Ginzburg-Landau theory.
The above values of $\alpha_{20}$ and $\gamma$ can be related
to $g_2/g_1$ and $g_3/g_1$ and they appear to be closer to the
second choice of $g_n$ used for Fig.~(\ref{Heat}).
The ratio of the two gaps,
as it changes along the $H_{c2}(T)$ line, is shown on the inset
in Fig.~\ref{Hc2}.
It varies from  $\Delta_1/\Delta_2\approx 2.3$ near $T_c=36$~K to
$\Delta_1/\Delta_2\approx 45$ at $T=18$~K, where the Ginzburg-Landau
theory breaks down. Due to a large difference in the $c$-axis
coherence lengths between the two bands, the smaller gap is quickly
suppressed by transverse magnetic field.
Also, the strong upward curvature of $H_{c2}(T)$ leads to temperature
variations of the anisotropy ratio $\gamma_{\rm an} = H_{c2}^\perp(T)/
H_{c2}^c(T)$, which changes from $\gamma_{\rm an}=1.7$ near $T_c$ to
$\gamma_{\rm an}=4.3$ at $T=18$~K. These values are again consistent
with experimental observations,\cite{cubitt1} as well as
with theoretical studies.\cite{miranovic,dahm,gurevich}

\subsection{Structure of a single vortex}

The structure of an isolated superconducting vortex parallel to the $c$-axis 
has been studied in MgB$_2$ by the scanning tunneling microscopy.\cite{eskildsen} 
Tunneling along the $c$-axis used in the experiment probes predominantly 
the three-dimensional $\pi$-band and the obtained spectra provide information
about a small passive gap. A large vortex core size of about 5 coherence 
lengths $\xi_c=\sqrt{\Phi_0/2\pi H^c_{c2}}$ was reported and attributed to 
a fast suppression of a passive gap by magnetic field, whereas the $c$-axis upper 
critical field is controlled by a large gap in the $\sigma$-band.\cite{eskildsen} 
The experimental observations were confirmed within the two-band model using the 
Bogoliubov-de Gennes\cite{nakai} and the Usadel equations.\cite{koshelev} 
We have, however, seen in the previous Subsection that a $\pi$-gap in MgB$_2$
is not suppressed near $H_{c2}(T)$ for fields applied along the $c$-axis.
To resolve this discrepancy we present here a systematic study of the vortex core 
in a two-gap superconductor in the framework of the Ginzburg-Landau theory.

\begin{figure}[t]
\begin{center}
\includegraphics[height=6cm,width=7.5cm]{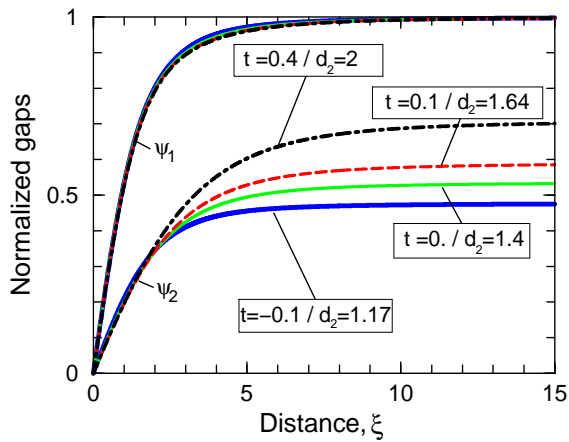}
\end{center}
\caption{Spatial dependencies of the gaps for various temperatures
with $t=\ln T_1/T\approx 1-T/T_1$ and $K_2/K_1=9$.}
\label{fig:vortT}
\end{figure}

We investigate the structure of a single-quantum vortex oriented 
parallel to the hexagonal $c$-axis. The two gaps are parametrized as
$\Delta_n(\mathbf{r})=\psi_n(r) e^{-i\theta}$, where $\theta$ is an
azimuthal angle and $r$ is a distance from the vortex axis. Since 
the Ginzburg-Landau parameter for MgB$_2$ is quite large,\cite{canfield}
$\kappa\simeq 25$, magnetic field can be neglected inside vortex core 
leading to the following system of the Ginzburg-Landau equations
\begin{equation}
\alpha_n \psi_n-\gamma \psi_{n'}+\beta_n \psi_n^3 -
K_n (\psi_n''+\psi_n'/r-Q^2\psi_n) = 0
\label{vortex}
\end{equation}
for $n=1,2$ ($n'=2,1$) and $Q\approx 1/r$. Away from the center of
a vortex, the two gaps approach their asymptotic amplitudes $\psi_{0n}$
\begin{equation}
\psi_{01} = \sqrt{\frac{a_1t+\gamma\rho}{\beta_1}}\ , \ \ \
\psi_{02} = \sqrt{\frac{\gamma/\rho-\alpha_2}{\beta_2}}
\end{equation}
with $\rho$ obeying Eq.~(\ref{ratio}). All distances are measured in
units of a temperature-dependent coherence length derived from
the upper critical field Eq.~(\ref{hc2_iso}). In order to solve
Eq.~(\ref{vortex}) numerically, a relaxation method has been used
\cite{numrecipes} on a linear array of 4000 points uniformly set
on a length of  $80\xi$ from the vortex center. An achieved accuracy
is of the order of $10^{-6}$.

The obtained results are shown in Figures 3--5, where amplitudes
$\psi_n(r)$ are normalized to the asymptotic value of the large gap
$\psi_{01}$. To quantify the size of the vortex core for each
component we determine the distance $d_n$, where $\psi_n(r)$ reaches
a half of its maximum value $\psi_{0n}$. In the case of a single-gap
superconductor such a distance is given within a few percent by the
coherence length. In a two-gap superconductor the characteristic
length scale for the large gap $d_1$ remains close to $\xi$,
while $d_2$ can substantially vary. Size of the vortex core
is given by $d_v={\rm max}(2d_1,2d_2)$.

\begin{figure}[t]
\begin{center}
\includegraphics[height=6cm,width=7.5cm]{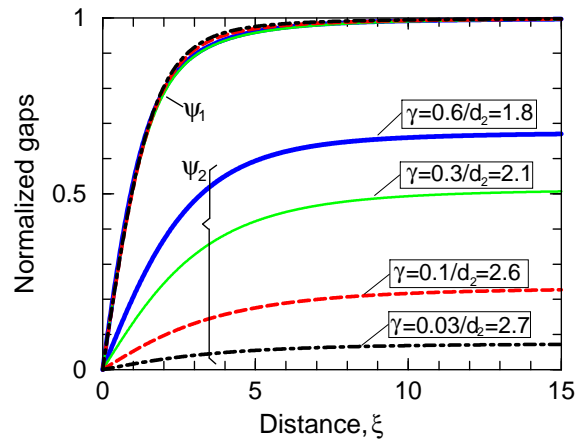}
\end{center}
\caption{Spatial dependencies of the gaps for various values of 
$\gamma$ (given in units of $a_1$) for $t=0.3$ and $K_2/K_1=9$.}
\label{fig:vortG}
\end{figure}

Results for temperature dependence of the vortex core are presented in 
Fig.~\ref{fig:vortT}. The parameters $\alpha_{20}$ and $\gamma$ are 
taken the same as in the study of the upper critical field, while we 
choose $K_2/K_1=9$ in order to amplify effect for the small gap. As was 
discussed above, the equilibrium ratio of the two gaps $\psi_{02}/\psi_{01}$ 
grows with decreasing temperature (increasing $t$). Simultaneously, the small 
gap becomes less constrained with its interaction to the large gap and the 
half-amplitude distance $d_2$ shows a noticeable growth. For $K_2\approx K_1$ 
such a less constrained behavior of $\psi_2(r)$ at low temperatures does not 
lead to an increase of the core size because both gaps have similar 
intrinsic coherence lengths.

This trend becomes more obvious if the coupling constant $\gamma$ is 
changed for fixed values of all other parameters, see Fig.~\ref{fig:vortG}.
For vanishing $\gamma$, the distance $d_2$ approaches asymptotically an 
intrinsic coherence length in the passive band. This length scale depends 
on $K_2$ ($d_2/d_1|_{\gamma=0}\simeq\sqrt{K_2/K_1}=3$), but is not directly  
related to an equilibrium value of the small gap: the small gap is 
reduced by a factor of 7 between $\gamma=0.6$ and $\gamma=0.03$, while 
the core size increases by 50\% only. Therefore, the single-band BCS 
estimate $\xi_2 = v_F/(\pi\Delta_2)$ for the characteristic length scale 
of the small gap sometimes used for interpretation of experimental 
data\cite{eskildsen} is not, in fact, applicable for a multi-gap 
superconductor.

\begin{figure}[t]
\begin{center}
\includegraphics[height=6cm,width=7.5cm]{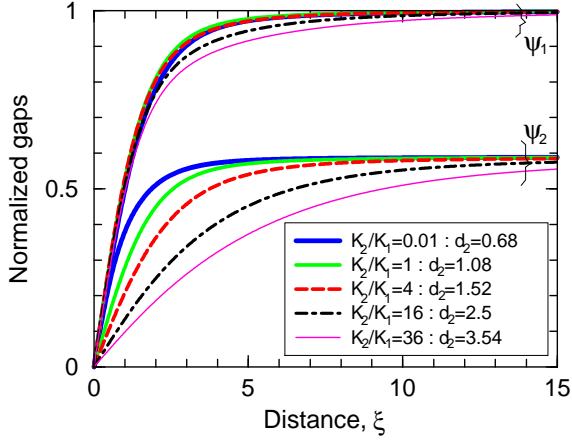}
\end{center}
\caption{Spatial dependencies of the gaps for various values of 
$K_2/K_1$ for $t=0.3$.}
\label{fig:vortK}
\end{figure}

Finally, Fig.~\ref{fig:vortK} presents evolution of the vortex 
core with varying ratio $K_2/K_1$, where again $\alpha_{20}/a_1=0.65$
and $\gamma/a_1=0.4$. The apparent size of the vortex core 
$d_v\simeq 2 d_2$ becomes about 5--6 coherence lengths for $K_2$ 
exceeding $K_1$ by an order of magnitude. For $K_2/K_1\simeq 1$, which 
follows from the band structure calculations, the vortex core size does 
not change significantly compared to the standard single-gap case. 
These results generally agree with the previous study,\cite{koshelev} 
though we conclude that unrealistically large values of $K_2/K_1$ are 
required to explain the experiment.\cite{eskildsen} Different strength 
of impurity scattering in the two bands cannot explain this discrepancy 
either. It is argued that the $\pi$-band is in the dirty limit.\cite{mazin}  
The coefficient $K_2$ in Eq.~(\ref{GL}) is accordingly replaced by a 
{\it smaller} diffusion constant. The numerical results (Fig.~\ref{fig:vortK})
as well as qualitative consideration show that in such a case the core size 
for $\psi_2(r)$ can only decrease. Note, that the spatial ansatz
$\psi(r)\sim\tanh(r/a)$ with $a=\xi$ used to fit the experimental 
data\cite{eskildsen} should be applied with $a=1.8\xi$ even for a single-gap
superconductor in the large $\kappa$ limit.\cite{hu}

\section{Orientation of vortex lattice}

Recent neutron scattering measurements\cite{cubitt2} in MgB$_2$ for 
fields along the hexagonal $c$-axis have discovered a new interesting
phase transition in the mixed state of this superconductor: 
a triangular vortex lattice rotates by 30$^\circ$ such that below 
the first transition field (0.5~T at $T=2$~K) a nearest-neighbor 
direction is aligned perpendicular to the crystal $a$-axis, whereas 
above the second transition field (0.9~T) the shortest intervortex 
spacing is parallel to the $a$-axis.\cite{cubitt2} We show in 
this section that such a peculiar behavior is determined by the 
two-gap nature of superconductivity in MgB$_2$.

\subsection{Single-gap superconductor}

An orientation of the flux line lattice in tetragonal and cubic 
superconductors has been theoretically studied by Takanaka.\cite{takanaka}
Recently, similar crystal field effects were found to be responsible 
for the formation of the square vortex lattices in the
borocarbides.\cite{dewilde,huse} The case of a single-gap 
hexagonal superconductor is treated by a straightforward generalization 
of the previous works. Symmetry arguments suggest that coupling between
the superconducting order parameter and a hexagonal crystal lattice
appears at the sixth-order gradient terms in the Ginzburg-Landau
functional. For simplicity, we assume that gap anisotropy is
negligible. Then, the six-order gradient terms derived from
the BCS theory are
\begin{eqnarray}
F_6 & = &
\frac{\zeta(7)N_0}{32\pi^6T_c^6}
\left(1-\frac{1}{2^7}\right)
\langle v_{Fi} v_{Fj}v_{Fk}v_{Fl}v_{Fm}v_{Fn}\rangle
\nonumber \\
& & \mbox{} \times
(\nabla_i\nabla_j\nabla_k\Delta^*)(\nabla_l\nabla_m\nabla_n\Delta) \ .
\end{eqnarray}
The above terms can be split into isotropic part and
anisotropic contribution, the latter being 
\begin{eqnarray}
F_6^{\rm an} & = & -\frac{\zeta(7)N_0}{64\pi^6T_c^6}
\left(1-\frac{1}{2^7}\right)
\bigl(\langle v_{Fx}^6\rangle - \langle v_{Fy}^6\rangle\bigr)
\nonumber \\
& \times & \Delta^*\Bigl[\nabla_x^6 -15\nabla_x^4\nabla_y^2
+ 15\nabla_x^2\nabla_y^4 -\nabla_y^6\Bigr]\Delta 
\label{F6} \\
& = &-\frac{1}{2}K_6 
\Delta^*\Bigl[(\nabla_x+i\nabla_y)^6+(\nabla_x-i\nabla_y)^6
\Bigr]\Delta \ . \nonumber
\end{eqnarray}
In this expression $\hat{\bf x}$ is fixed to the $a$-axis in the 
basal plane. (An alternative choice is the $b$-axis.) 
If $\hat{\bf x}$ and $\hat{\bf y}$ are simultaneously rotated by 
angle $\varphi$ about the $c$-axis, $(\nabla_x\pm i\nabla_y)^6$
acquires an extra factor $e^{\pm 6i\varphi}$.
In the following we always make such a rotation in order to have 
$\hat{\bf x}$ pointing between nearest-neighbor vortices. 
Periodic Abrikosov solutions with chains of vortices parallel
to the $x$-axis are most easily 
written in the Landau gauge ${\bf A}=(-Hy,0,0)$.\cite{SaintJ}

The higher order gradient terms Eq.~(\ref{F6}) give a small
factor $H^2\sim (1-T/T_c)^2$ and can be treated as a
perturbation in the Ginzburg-Landau regime. The Landau levels expansion
yields $\Delta(x)= c_0 f_0(x) + c_6 f_6(x)+...$, where the coefficient
for the admixed sixth Landau level is
$c_6/c_0\approx-(\sqrt{6!}/3)h^2e^{6i\varphi}K_6/K$. When substituted
into the quartic Ginzburg-Landau term, this expression produces
the following angular dependent part of the free energy:
\begin{equation}
\delta F(\varphi)= - \frac{2\sqrt{6!}K_6}{3K}h_{c2}^2 \beta |c_0|^4
\langle |f_0|^2f^*_0f_6\rangle
\cos(6\varphi)\ ,
\label{Fan}
\end{equation}
with $|c_0|^2 = K(h_{c2}-h)\langle |f_0|^2\rangle/
(\beta\langle |f_0|^4\rangle)$. Spatial averaging of the combination 
of the Landau levels is done in a standard way
\begin{eqnarray}
\frac{\langle |f_0|^2f^*_0f_6\rangle}{\langle |f_0|^2\rangle^2} & = & 
\frac{\sqrt{\sigma}}{12\sqrt{5}} \sum_{n,m} \cos(2\pi\rho nm)
e^{-\pi\sigma(n^2+m^2)} \nonumber \\
&&\mbox{}\times\Bigl[\pi^3\sigma^3(n-m)^6-\frac{15}{2}\pi^2\sigma^2(n-m)^4
\nonumber \\
&& \mbox{}\ \ \ +\frac{45}{4}\pi\sigma(n-m)^2-\frac{15}{8}
\Bigr]\  ,
\end{eqnarray}
where summation goes over all integer $n$ and $m$ and parameters
$\rho$ and $\sigma$ describe an arbitrary vortex lattice.\cite{SaintJ}
For a hexagonal lattice ($\rho=1/2$, $\sigma=\sqrt{3}/2$), 
the numerical value of the lattice factor is
$\langle|f_0|^2f^*_0f_6\rangle/\langle |f_0|^4\rangle=
-0.279$. 
Hence, $\delta F(\varphi) \simeq +K_6\cos(6\varphi)$ and
for $\langle v_{Fx}^6\rangle > \langle v_{Fy}^6\rangle$
($K_6>0$) 
the equilibrium angle is $\varphi=\pi/2$ ($\pi/6$), which means that 
the shortest spacing between vortices in a triangular lattice is oriented 
perpendicular to the $a$-axis, while for the other sign of
anisotropy the shortest side of a vortex triangle is along the 
$a$-axis. Thus, the Fermi surface anisotropy fixes uniquely 
the orientation of the flux line lattice near the upper critical field.

\subsection{Two-gap superconductor}

In a multiband superconductor effect of crystal anisotropy may vary 
from one sheet of the Fermi surface to another. We apply again the
tight-binding representation \cite{kong} to obtain a quantitative 
insight about such effects in MgB$_2$. Explicit expressions for
dispersions of the two hole $\sigma$-bands are presented in the Appendix.
Hexagonal anisotropy in the narrow $\sigma$-cylinders is enhanced by 
a nonanalytic form of the hole dispersions. Combined anisotropy of 
the $\sigma$-band is $\langle v_{Fx}^6\rangle = 4.608$, 
$\langle v_{Fy}^6\rangle = 4.601$, while for the $\pi$-band
$\langle v_{Fx}^6\rangle = 1.514$, $\langle v_{Fy}^6\rangle = 1.776$
in units of 10$^{46}$~(cm/s)$^6$. 
According to the choice of the coordinate system,\cite{kong} the
$\hat{x}$-axis is parallel to the $b$-direction and the $\hat{y}$-axis 
is parallel to the $a$-direction in the boron plane.
The above values might be not very accurate due to uncertainty of the LDA 
results, however, they suggest two special qualitative features for 
MgB$_2$. First, relative hexagonal anisotropy of the Fermi velocity 
$v_{Fn}(\varphi)$ differs by almost two orders of magnitude between 
the two sets of bands. Second, corresponding hexagonal terms
have different signs in the two bands.
In Appendix, we have shown that the sign difference is a robust
feature of the tight-binding approximation and cannot be changed by
a small change of the tight-binding parameters.

We investigate equilibrium orientation of the vortex lattice
in MgB$_2$ within the two-gap Ginzburg-Landau theory. Anisotropic 
sixth-order gradient terms of the type (\ref{F6}) have to be added 
to the functional (\ref{GL}) separately for each of the two 
superconducting order parameters. As was discussed in the previous 
paragraph the anisotropy constants have different signs $K_{61}>0$ 
and $K_{62}<0$ and obey $|K_{61}|\ll |K_{62}|$. In the vicinity of 
the upper critical field the two gaps are expanded as 
$\Delta_1(x)= c_0 f_0(x) + c_6 f_6(x)$ and 
$\Delta_2(x)= d_0 f_0(x) + d_6 f_6(x)$. Solution of the linearized 
Ginzburg-Landau equations yields the following amplitudes for 
the sixth Landau levels:
\begin{eqnarray}
c_6 & = & - 4\sqrt{6!}h^3 e^{6i\varphi} \frac{K_{61}\tilde{\alpha}_2c_0
+ K_{62}\gamma d_0}{\tilde{\alpha}_1\tilde{\alpha}_2-\gamma^2} \ ,
\nonumber \\
d_6 & = & - 4\sqrt{6!}h^3 e^{6i\varphi} \frac{K_{62}\tilde{\alpha}_1d_0
+ K_{61}\gamma c_0}{\tilde{\alpha}_1\tilde{\alpha}_2-\gamma^2} 
\label{coeffs}
\end{eqnarray}
with $\tilde{\alpha}_{1,2}=\alpha_{1,2} + 13K_{1,2}h$. Subsequent 
calculations follow closely the single-gap case from the preceding 
subsection. The angular dependent part of the free energy is obtained
by substituting (\ref{coeffs}) into the fourth-order terms:
\begin{equation}
\delta F(\varphi) = \Bigl[\beta_1 c_0^3(c_6+c_6^*) +
\beta_2 d_0^3(d_6+d_6^*)\Bigr]\langle |f_0|^2f^*_0f_6\rangle \ .
\label{Fan2a}
\end{equation}
The resulting expression can be greatly simplified if one uses
$(\Delta_2/\Delta_1)^2\simeq 0.1$ as a small parameter.
With accuracy $O[(\Delta_2/\Delta_1)^4]$ we can neglect 
the angular dependent part determined by the small gap. This yields
in a close analogy with Eq.~(\ref{Fan}) the following anisotropy
energy for the vortex lattice near $H_{c2}$
\begin{eqnarray}
&& \delta F(\varphi)= -\frac{2\sqrt{6!}}{3K_1}h_{c2}^2 \beta_1 |c_0|^4
\langle |f_0|^2f^*_0f_6\rangle K_6^{\rm eff}\cos(6\varphi) \ ,
\nonumber \\
&&  K_6^{\rm eff} = K_{61} + K_{62}
\frac{\gamma^2}{(\alpha_2 +K_2h)(\alpha_2+13K_2h)} \ .
\label{Fan2b}
\end{eqnarray}
Despite the fact that we have omitted terms $\sim d_0^3d_6$, the Fermi 
surface anisotropy of the second band still contributes to the effective 
anisotropy constant $K_6^{\rm eff}$
via linearized Ginzburg-Landau 
equations. Along the upper critical line this contribution decreases 
suggesting
the following scenario for MgB$_2$. 

In the region near $T_c$
the second band makes the largest contribution
to $K_6^{\rm eff}$: a small factor $\gamma^2/\alpha_2^2\sim 0.1$
is outweighed by $|K_{61}/K_{62}|<0.1$. As a result, $K_6^{\rm eff}$
is negative  and $\varphi=0$, which means that the shortest intervortex
spacing is parallel to the $b$-axis. At lower temperatures
and higher magnetic fields the second term in $K_6^{\rm eff}$
decreases and the Fermi surface anisotropy of the first band starts 
to determine the (positive) sign of $K_6^{\rm eff}$. In this case,
$\varphi=\pi/2$ ($\pi/6$) and the side of the vortex triangle is parallel
to the $a$-axis. The very small $|K_{61}/K_{62}|=1.8\cdot 10^{-2}$, which
follows from the band structure data,\cite{kong} is
insufficient to have such a reorientation transition in the Ginzburg-Landau
region. Absolute values of anisotropy coefficients are, however,
quite sensitive to the precise values of the tight-binding parameters
and it is reasonable to assume that experimental values of $K_{6n}$ 
are such that the reorientation transition is allowed.

The derived sequence of the orientations of the flux line lattice in
MgB$_2$ completely agrees with the neutron scattering data,\cite{cubitt2}
though we have used a different scan line in the $H$--$T$ plane in order 
to to demonstrate the presence of the $30^\circ$-orientational transformation, 
see Fig.~6. Condition $K_6^{\rm eff}=0$, or similar one
applied to Eq.~(\ref{Fan2a}), defines a line $H^*(T)$ in the $H$--$T$
plane, which has a negative slope at the crossing point with $H_{c2}(T)$. 
The six-fold anisotropy for the vortex lattice vanishes along $H^*(T)$ 
and all orientations with different angles $\varphi$ become degenerate 
in the adopted approximation. The sequence of orientational 
phase transition in such a case depends on weaker 
higher-order harmonics. One can generally write
\begin{equation}
\delta F(\varphi) = K_6 \cos(6\varphi) + K_{12} \cos(12\varphi) \ ,
\label{deltaF}
\end{equation}
where the higher-order harmonics comes with a small
coefficient $|K_{12}|\ll |K_6|$. Depending on the sign of $K_{12}$
transformation between low-field $\varphi=0$ and high-field 
$\varphi=\pi/6$ ($\pi/2$) orientations, when $K_6$ changes sign, goes 
either via two second-order transitions ($K_{12}>0$) or via single 
first-order transition ($K_{12}<0$). In the former case the transitions take 
place at $K_6=\pm 4K_{12}$, whereas in the latter case the first-order 
transition is at $K_6=0$. These conclusions are easily obtained by comparing 
the energy of a saddle-point solution $\cos(6\varphi)=-K_6/(4K_{12})$
for Eq.~(\ref{deltaF}), which is $\delta F_{sp} = - K_6^2/(8K_{12})$,
to the energies of two extreme orientations.

\begin{figure}[t]
\begin{center}
\includegraphics[width=0.8\columnwidth]{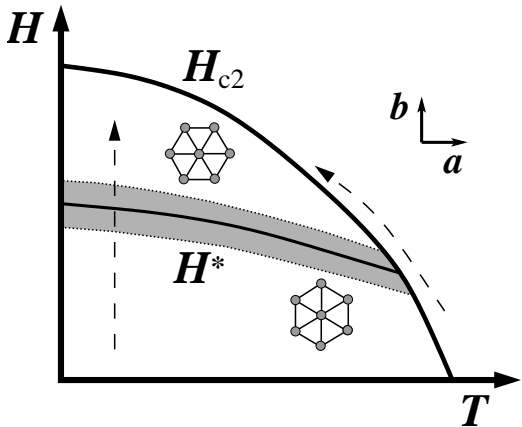}
\end{center}
\caption{
Phase diagram of MgB$_2$ for fields parallel to the $c$-axis.
Shaded region corresponds to intermediate orientations of the vortex 
lattice separated by dotted lines of the second-order transitions.
Dashed lines indicate the scans used in the experiment (vertical) and
in the presented theory.}
\end{figure}

In order to determine sign of the higher-order harmonics for a
two-gap superconductor we expand the fourth-order terms in 
the Ginzburg-Landau functional (\ref{GL}) to the next order
\begin{equation}
\delta F'(\varphi) = \frac{1}{2}\Bigl[\beta_1 c_0^2(c_6^2+c_6^{*2}) +
\beta_2 d_0^2(d_6^2+d_6^{*2})\Bigr]\langle f_0^{*2}f_6^2\rangle \ .
\label{Fan3a}
\end{equation}
These terms are responsible for a $\cos(12\varphi)$ anisotropy introduced 
before. Similar angular dependence is also induced by higher-order harmonics 
of the Fermi velocity $v_F(\varphi)$, though our estimate shows that even for 
the $\pi$-bands corresponding modulations are very small.\cite{Lasher}
Sign of $\cos(12\varphi)$ term in Eq.~(\ref{Fan3a}) 
depends only on a geometric factor, spatial average of
the Landau levels wave functions. We find for a perfect triangular
lattice $\langle f_0^{*2}f_6^2\rangle/\langle |f_0|^2\rangle^2=0.804$. 
Thus, the twelveth-order harmonics in Eq.~(\ref{deltaF}) has a positive 
coefficient and transformation between the low-field state with $\varphi=0$ 
and the high-field state $\varphi=\pi/6$ goes via a phase with intermediate 
values of $\varphi$ separated by two second order transitions.

The anisotropy terms of the type (\ref{F6}) also produce a six-fold
modulation of the upper critical field in the basal plane.
Sign of the corresponding modulations of $H_{c2}(\varphi)$ should
also change at a certain temperature, which is determined 
by a suppression of the small gap in transverse magnetic field and is not, 
therefore, related to the intersection point of $H_{c2}(T)$ and $H^*(T)$ 
lines on the phase diagram for $H\parallel c$, Fig.~6.

\section{Conclusions}

We have derived the Ginzburg-Landau functional of a two-gap
superconductor within the weak-coupling BCS theory.
The functional contains only a single interaction
term between the two superconducting gaps (condensates).
This property allows a meaningful analysis of various
magnetic properties of a multi-gap superconductor in
the framework of the Ginzburg-Landau theory.
Apart from confirming the previous results on an unusual 
temperature dependence of the transverse upper critical field in
MgB$_2$, we have presented detailed
investigation of the vortex core structure and have shown
that the orientational phase transitions observed 
in the flux line lattice in MgB$_2$ is a manifestation of the multi-band
nature of superconductivity in this material.
The proposed minimal model for the $30^\circ$-rotation of the
vortex lattice includes only anisotropy of the Fermi surface.
An additional source of  six-fold anisotropy for the vortex
lattice can arise from angular dependence of the superconducting
gap. It was argued that the latter source of (four-fold) anisotropy
is essential for physics of the square to distorted triangular
lattice transition in the mixed state of borocarbides.\cite{gap_anis}
For phonon-mediated superconductivity in MgB$_2$, the gap modulations
should be quite small, especially for the large gap on the narrow
$\sigma$-cylinders of the Fermi surface. Experimentally, the role
of gap anisotropy can be judged from the position of $H^*(T)$ line
in the $H$--$T$ plane. $H^*(T)$ does not cross $H_{c2}(T)$ line
in scenarios with significant gap anisotropy.\cite{gap_anis}
A further insight in anisotropic properties of different Fermi surface
sheets in MgB$_2$ can be obtained by studying experimentally and theoretically 
the hexagonal anisotropy of the upper critical field in the basal plane.

\begin{acknowledgments}
The authors would like to acknowledge useful discussions with R. Cubitt, 
M. R. Eskildsen, V. M. Gvozdikov, A. G. M. Jansen, S. M. Kazakov, K. Machida, 
I. I. Mazin, and  V. P. Mineev. We also thank F. Bouquet and P. Samuely for 
providing their experimental data.
\end{acknowledgments}

\appendix*
\section{Anisotropy in $\sigma$-bands}

We give here expressions for the dispersions and Fermi surface
anisotropies in the two $\sigma$-bands, which
are derived from the tight-binding fits of
Kong {\it et al\/}.\cite{kong} The in-plane $p_{x,y}$ boron orbitals
in MgB$_2$ undergo an $sp^2$-hybridization
with $s$-orbitals and form three bonding bands.
At ${\bf k}_\perp=0$ these
bands are split into a nondegenerate $A$-symmetric band
and doubly-degenerate $E$-symmetric band, which lies slightly
above the Fermi level. Away from the ${\bf k}_\perp=0$-line
the $E$-band splits into light and heavy hole bands. Their dispersions
obtained by expansion of the tight-binding matrix\cite{kong} in small
$k_\perp$ are
\begin{eqnarray*}
\varepsilon_l({\bf k})& = & \varepsilon(k_z) - 2t_\perp\biggl[
{\textstyle\frac{1}{8}}k_\perp^2 + d k_\perp^4
\frac{2g({\bf k})+1-1/d}{384(1+d)}\biggr],\\
\varepsilon_h({\bf k})& = & \varepsilon(k_z) - 2t_\perp\biggl[
{\textstyle\frac{3}{8}}dk_\perp^2 - d k_\perp^4
\frac{2g({\bf k})+ 7 + 9d}{384(1+d)}\biggr],
\end{eqnarray*}
where $\varepsilon(k_z) = \varepsilon_0 - 2t_z \cos k_z$ and
$g({\bf k}) = (k_x^6-15k_x^4k_y^2 +15k_x^2k_y^4-k_y^6)/k_\perp^6$.
The tight-binding parameters presented in Ref.~\onlinecite{kong}
are $\varepsilon_0=0.58$~eV, $t_\perp=5.69$~eV,
$t_z=0.094$~eV, and $d=0.16$. The six-fold anisotropy
is given by unusual nonanalytic terms, which are formally
of the fourth order in $k$. Appearance of such nonanalytic
terms is a direct consequence of the degeneracy of the two bands
at $k=0$.
For example, a nonanalytic form of $\varepsilon({\bf k})$ is known
for four-fold degenerate hole bands of Si and Ge,\cite{kittel}
which have cubic anisotropy already in $O(k^2)$ order.
Nonanalyticity of $\varepsilon_{l,h}({\bf k})$ leads to a
relative enhancement of the hexagonal anisotropy on
two narrow Fermi surface cylinders.
This anisotropy has opposite sign in light- and heavy-hole
bands. The net anisotropy of the combined $\sigma$-band is
determined mostly by the light-holes, which have larger
in-plane Fermi velocities.

\end{document}